\documentstyle[pra,aps,psfig,subfigure]{revtex}
\def \greatersim{\mathrel{\lower1pt\hbox{$>\atop\raise2pt\hbox{$\sim$}$}}}
\begin{document}
\draft
\twocolumn[\hsize\textwidth\columnwidth\hsize\csname @twocolumnfalse\endcsname
\title{Nucleation of vortex arrays in rotating anisotropic Bose-Einstein
condensates}
\author{David L. Feder,$^{1,2}$ Charles W. Clark,$^2$ and Barry I.
Schneider$^3$}
\address{$^1$University of Oxford, Parks Road, Oxford OX1 3PU, U.K.}
\address{$^2$National Institute of Standards and Technology, Gaithersburg, MD
20899-8410}
\address{$^3$Physics Division, National Science Foundation, Arlington,
Virginia 22230}
\date{\today}
\maketitle
\begin{abstract}
The nucleation of vortices and the resulting structures of vortex arrays in
dilute, trapped, zero-temperature Bose-Einstein condensates are
investigated numerically. Vortices are generated by rotating a
three-dimensional, anisotropic harmonic atom trap. The condensate
ground state is obtained by propagating the Gross-Pitaevskii equation
in imaginary time.  Vortices first appear at a rotation frequency
significantly larger than the critical frequency for vortex stabilization.
This is consistent with a critical velocity mechanism for vortex
nucleation. At higher frequencies, the structures of the vortex arrays are
strongly influenced by trap geometry.
\end{abstract}
]

\narrowtext

Since the experimental achievement of Bose-Einstein condensation (BEC) in
confined alkali gases~\cite{Cornell,Ketterle,Hulet,Others}, the possibility of
generating vortices in confined weakly-interacting dilute Bose gases has been
intensively
studied~\cite{McCann,Adams,Dum,Marshall,Davies,Marzlin,Holland,Cornell2}.
While theoretical investigations of stability have generally been restricted
to the case of a single vortex~\cite{Dalfovo,Fetter1,Muller,Pu,Feder1}, the
proposed experimental techniques may induce several vortices
simultaneously~\cite{Rokhsar}. Under appropriate stabilizing conditions,
such as a continuously applied torque, these vortices would form an array akin
to those obtained in rotating superfluid
helium~\cite{Packard1,Tkachenko,Campbell}.

A standard approach used to `spin-up' superfluid helium is to rotate the
container at an angular frequency $\Omega$. Aside from significant hysteresis
effects~\cite{Packard2,Jones}, vortices tend to first appear at a frequency
$\Omega_{\nu}$, whose value is comparable to the critical frequency $\Omega_c$
at which the presence of vortices lowers the free energy of the interacting
system~\cite{Fetter2}. Energy minimization arguments have also yielded vortex
arrays that are very similar to those observed
experimentally~\cite{Tkachenko,Campbell}. Despite these successes, the
mechanisms for vortex {\it nucleation} by rotation remain poorly understood;
important factors are thought to include presence of
a normal fluid, impurities, and surface roughness~\cite{Jones,Donnely,Aranson}.

It has been suggested that vortices may be similarly generated in the dilute
Bose gases by rotating the trap about its center~\cite{Fetter1,Rokhsar}.
Evidently, a harmonic potential can transfer angular momentum to the
gas only if it is anisotropic in the plane of rotation. While vortices in
such a system at zero temperature have been shown to become energetically
stable for $\Omega>\Omega_c$~\cite{Dalfovo,Fetter1,Feder1}, the particle flow
could remain irrotational at these angular frequencies since there exists an
energy barrier to vortex formation~\cite{Fetter1}. Suppression of this barrier
could be induced by application of a perturbing potential near the edge of the
confined gas, as has been simulated in the low-density limit~\cite{Rokhsar}.
One of the primary motivations for the present work, however, is to determine
if there exists any intrinsic mechanism for vortex nucleation in a dilute
quantum fluid that is free of impurities, surface effects, and thermal atoms.
We find that vortices can indeed be generated by rotating Bose condensates
confined in an anisotropic harmonic trap. The value of $\Omega_{\nu}$ at which
vortices are spontaneously nucleated is somewhat larger than $\Omega_c$. For
$\Omega>\Omega_{\nu}$ multiple vortices appear simultaneously, in patterns
that depend upon the geometry of the trap.

The dynamics of a dilute Bose condensate at zero temperature are
governed by the time-dependent Gross-Pitaevskii (GP) equation~\cite{GP}.
Previous simulations of the GP equation have demonstrated that
vortex-antivortex pairs or vortex half-rings can be generated by superflow
around a stationary obstacle~\cite{McCann,Adams,Davies,Frisch} or through a
small aperture~\cite{Burkhart}. In \onlinecite{McCann}, the vortex pairs
form when the magnitude of the superfluid velocity exceeds a critical value
which is proportional to the local sound velocity; recent experimental results
support this conclusion~\cite{Ketterle2}. To our knowledge, no numerical
investigation of vortex nucleation in three-dimensional inhomogeneous rotating
superfluids has hitherto been attempted.

The numerical calculations presented here model the experimental apparatus of 
Kozuma {\it et al.}~\cite{Phillips}, where $^{23}$Na atoms are confined in a
completely anisotropic three-dimensional harmonic oscillator potential.
In the presence of a constant external torque, the condensate obeys the
time-dependent GP equation in the rotating reference frame:

\begin{equation}
i\partial_{\tau}\psi({\bf r},\tau)
=\left[T+V_{\rm trap}+V_{\rm H}-\Omega L_z\right]\psi({\bf r},\tau),
\label{gp}
\end{equation}

\noindent where the kinetic energy is $T=-{\case1/2}\vec{\nabla}^2$, the trap
potential is
$V_{\rm trap}={\case1/2}\left(x^2+\alpha^2y^2+\beta^2z^2\right)$, and the
Hartree term is $V_{\rm H}=4\pi\eta|\psi|^2$. The angular momentum operator 
$L_z=i\left(y\partial_x-x\partial_y\right)$ rotates the system about the
$z$-axis at the trap center at a constant angular frequency $\Omega$. The
trapping frequencies are
$(\omega_x,\omega_y,\omega_z)=\omega_x(1,\alpha,\beta)$ with
$\omega_x=2\pi\times 26.87$~rad/s, $\alpha=\sqrt{2}$, and $\beta=1/\sqrt{2}$.
Normalizing the condensate $\int d{\bf r}|\psi({\bf r},\tau)|^2=1$ yields the
scaling parameter $\eta=N_0a/d_x$, where $a=2.75$~nm is the s-wave scattering
length for Na and $N_0$ is the number of condensate atoms. Unless explicitly
written, energy, length, and time are given throughout in scaled harmonic
oscillator units $\hbar\omega_x$, $d_x=\sqrt{\hbar/M\omega_x}\approx 4.0~\mu$m,
and ${\rm T}=\omega_x^{-1}\approx 6$~ms, respectively.

The stationary ground-state solution of the GP equation, defined as that which
minimizes the value of the chemical potential, is found by norm-preserving
imaginary time propagation (the method of steepest descents) using an adaptive
stepsize Runge-Kutta integrator. The complex condensate wavefunction is
expressed within a discrete-variable representation (DVR)~\cite{Feder2} based
on Gauss-Hermite quadrature, and is assumed to be even under inversion of $z$.
The numerical techniques are described in greater detail
elsewhere~\cite{Feder1,Feder2}. The initial state (at zero imaginary
time $\tilde{\tau}\equiv i\tau=0$) is taken to be the vortex-free Thomas-Fermi
(TF) wavefunction $\psi_{\rm TF}=\sqrt{(\mu_{\rm TF}-V_{\rm trap})/4\pi\eta}$,
which is the time-independent solution of Eq.~(\ref{gp}), neglecting $T$ and
$L_z$, with chemical potential
$\mu_{\rm TF}={\case1/2}(15\alpha\beta\eta)^{2/5}$. The GP equation for a
given value of $\Omega$ and $N_0$ is propagated in imaginary time until the
fluctuations in both the chemical potential and the norm become smaller than
$10^{-11}$. It should be emphasized that the equilibrium configuration is
found not to depend on the choice of purely real initial state. Since the 
final state is unconstrained except for $z$-parity, the lowest-lying
eigenfunction of the GP equation corresponds to a local minimum of the free
energy functional.

\begin{figure}
\psfig{figure=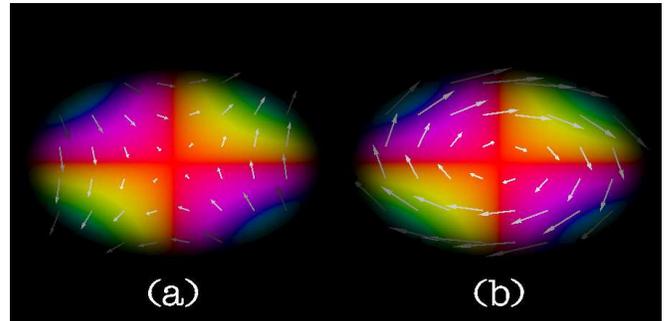,width=\columnwidth,angle=0}
\caption{The condensate density integrated down the axis of rotation
(brightness) and the phase $\varphi$ through the $z=0$ plane (colors in the
sequence red-green-blue-red for $\varphi=0$ through $2\pi$) are shown for
$N_0=10^6$ and an angular frequency $\Omega=0.45\omega_x$ applied
counterclockwise. The irrotational velocity field in the laboratory (a) and
rotating (b) frames of reference are shown as arrows, whose lengths are
proportional to the magnitude of the velocity $|{\bf v}_s|$ at the quiver tail.}
\label{irrot}
\end{figure}

In Fig.~\ref{irrot} are depicted the condensate density, which is stationary
in the rotating frame, as well as the condensate phase and the velocity field
in the laboratory and rotating frames, for $\Omega=0.45\omega_x$ and
$N_0=10^6$. The density profile at this angular frequency contains no vortices
but is slightly extended from that of a non-rotating condensate due to the
centrifugal forces. The velocity field in the laboratory frame is given by
${\bf v}_s^l\equiv\vec{\nabla}\varphi$ in units of $\omega_xd_x$, where
$\varphi$ is the condensate phase. In the rotating frame,
${\bf v}_s^r={\bf v}_s^l-\Omega\hat{z}\times{\bf r}$. There are no closed
velocity streamlines found in Fig.~\ref{irrot}(a). Such an irrotational flow
$\vec{\nabla}\times{\bf v}_s=0$ is characteristic of a superfluid, distinct
from the related properties of vortex quantization and stability. The only
solution of the GP equation satisfying irrotational flow in a
cylindrically-symmetric trap is ${\bf v}_s=0$: rotating the trap is equivalent
to doing nothing. The irrotational velocity field for an anisotropic trap is
nontrivial, however. Since the density profile is independent of orientation,
mass flow must accompany the rotation even though the superfluid prefers to
remain at rest~\cite{Fetter2}.

The condensate is found to remain vortex-free for angular velocities
significantly larger than the expected critical frequency for the stability of
a single vortex $\Omega_c^{(1)}$~\cite{Dalfovo,Fetter1,Feder1}. In order to
determine if irrotational configurations correspond to the global free energy
minima of the system, vortex states are investigated by artificially imposing
total circulation $n\kappa$ on the condensate wavefunction. By winding the
phase at $\tilde{\tau}=0$ by $2\pi n$ about the trap center, imaginary-time
propagation of the GP equation yields the minimum energy configuration with
$n$ vortices if such a solution is stationary or metastable~\cite{Feder1}.
The results for $N_0=10^6$ and $\Omega=0.45\omega_x$ are summarized in
Table~\ref{tablerot}. At this angular frequency, states with $n=1,2,3$ are all
energetically favored over the vortex-free solution. The vortices in these
cases are predominantly oriented along the ($\hat{z}$) axis of rotation, and
are located symmetrically about the origin on the (loose) $x$-axis. The
frequency chosen is too low to support the four vortex case
$\Omega<\Omega_c^{(4)}$, but is larger than the frequency (which may correspond
to metastability) at which the chemical potentials for $n=3$ and $n=4$ cross.

\begin{figure}[tb]
\centering
\subfigure{\psfig{figure=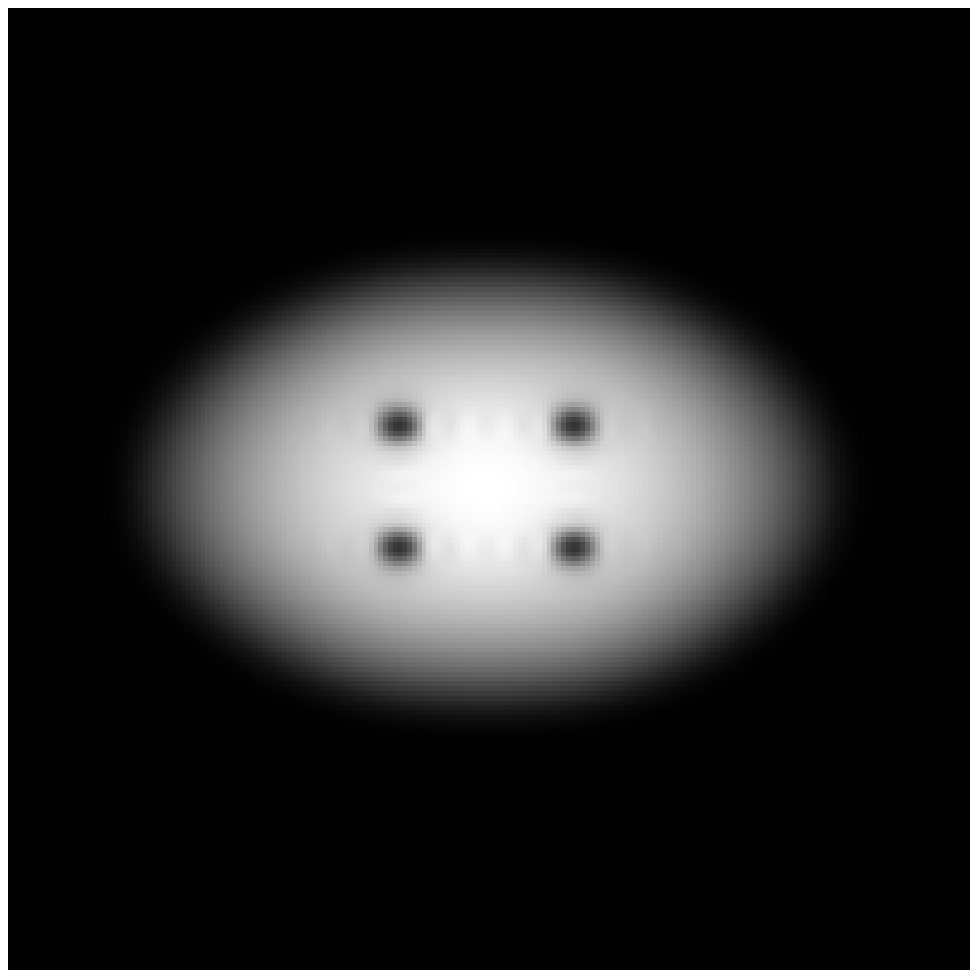,width=0.49\columnwidth,angle=90}}
\hspace{-0.13in}
\subfigure{\psfig{figure=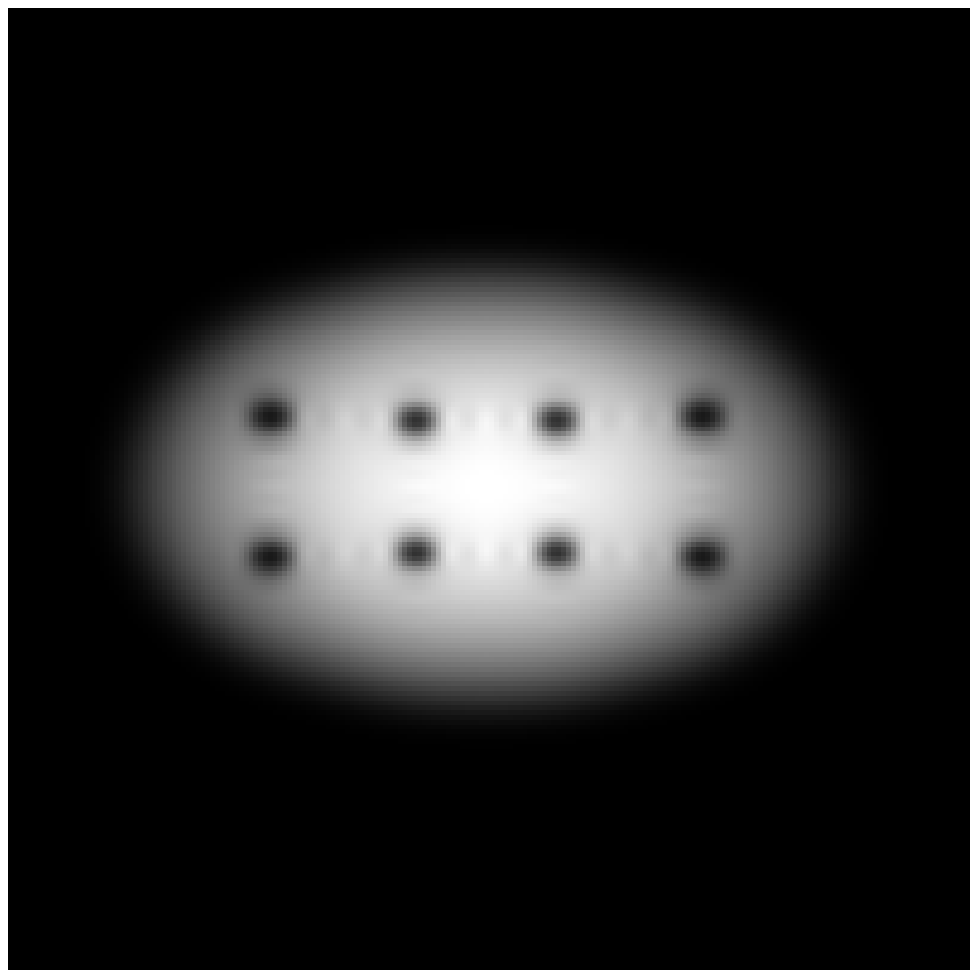,width=0.49\columnwidth,angle=90}}\\
\vspace{-0.75in}
\subfigure{\psfig{figure=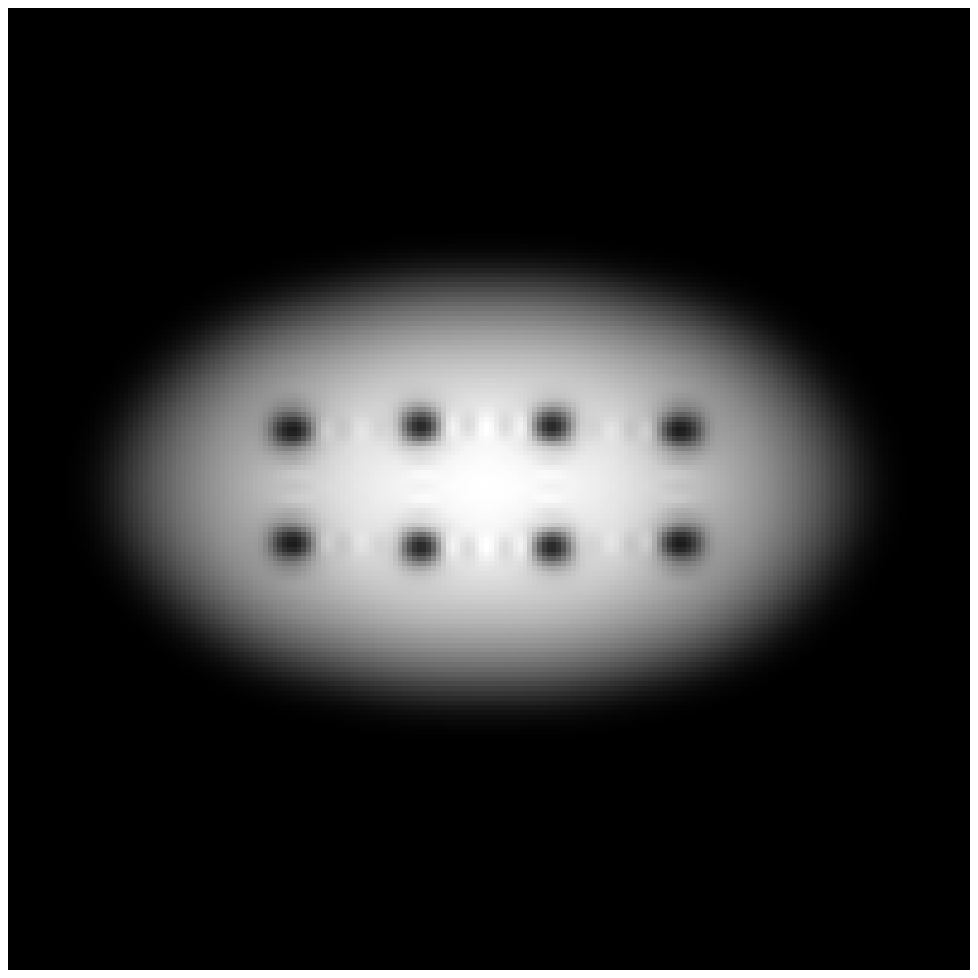,width=0.49\columnwidth,angle=90}}
\hspace{-0.13in}
\subfigure{\psfig{figure=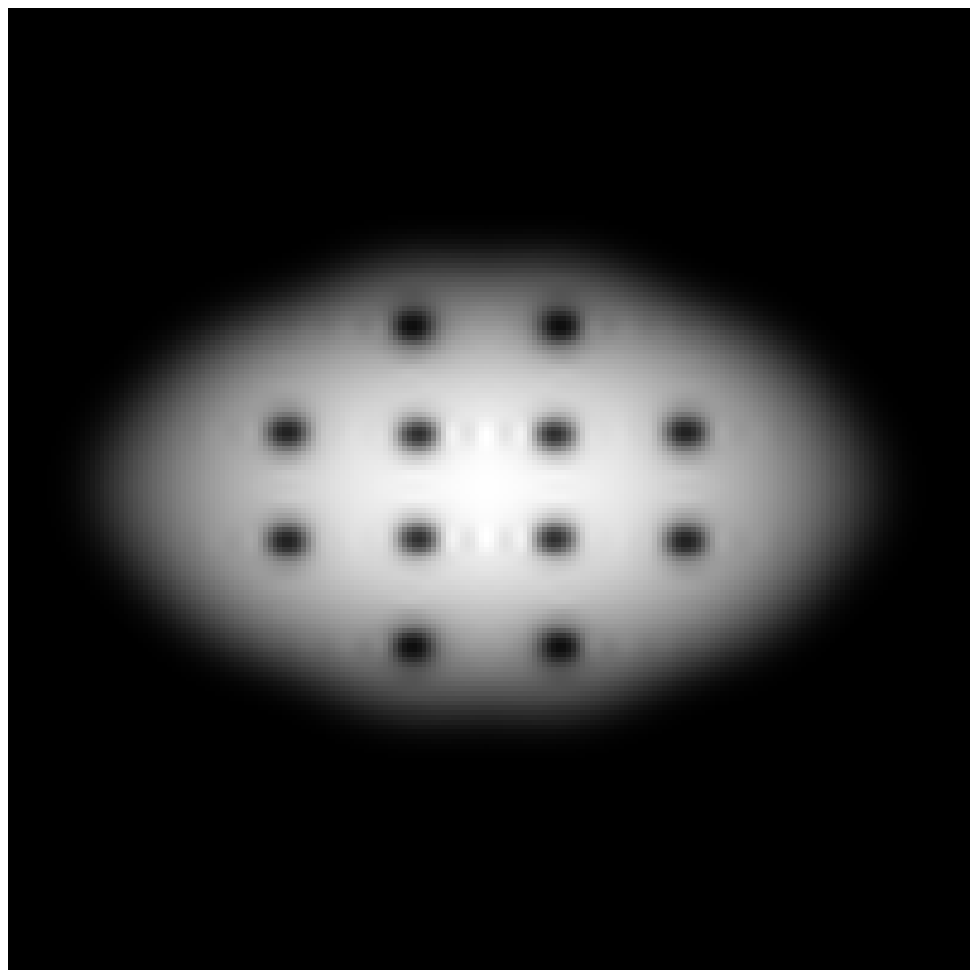,width=0.49\columnwidth,angle=90}}\\
\vspace{-0.75in}
\subfigure{\psfig{figure=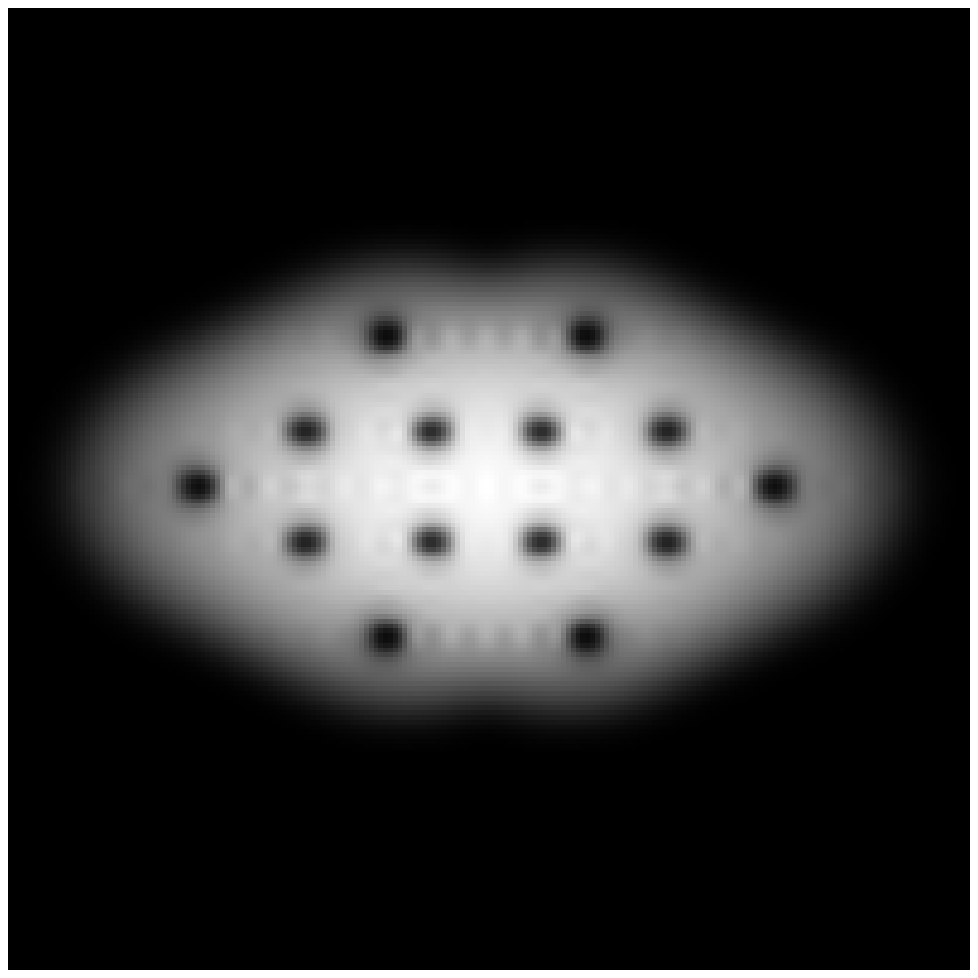,width=0.49\columnwidth,angle=90}}
\hspace{-0.13in}
\subfigure{\psfig{figure=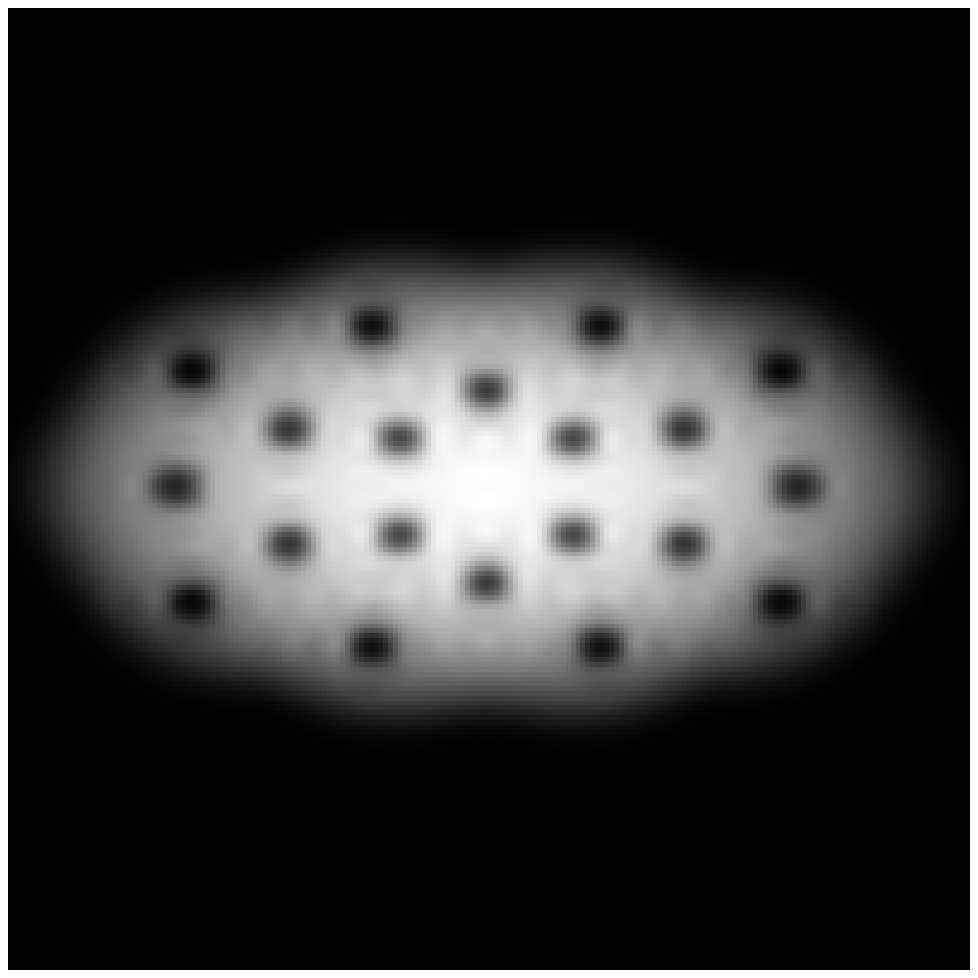,width=0.49\columnwidth,angle=90}}
\caption{The condensate densities integrated down the axis of rotation
(perpendicular to the page) are shown in the rotating frame as a function of
angular velocity $\Omega$ for $N_0=10^6$. From the top left to bottom right in
raster order, are shown $\Omega=0.5\omega_x$, $0.55\omega_x$, $0.65\omega_x$,
$0.7\omega_x$, $0.75\omega_x$, and $0.8\omega_x$. The density is proportional
to the brightness, and the dark spots correspond to unit vortices. Each box is
$18d_x\approx 73\mu$m wide.}
\label{rot}
\end{figure}

As shown in Fig.~\ref{rot}, vortices with the same circulation $\kappa$ (as
opposed to vortex-antivortex pairs) begin to penetrate the cloud above a
critical angular velocity for vortex nucleation $\Omega_{\nu}$. The value of
$\Omega_{\nu}$ is found to not depend strongly on trap geometry and to
decrease very slowly with $N_0$; for $N_0=10^q$ with $q=\{5,6,7\}$, we obtain
$\Omega_{\nu}=\{0.65,\,0.50,\,0.36\}\omega_x\pm 0.01\omega_x$, respectively.
In contrast, the critical frequency for the stabilization of a single vortex
in an anisotropic trap is approximately given by the TF expression
$\Omega_c^{(1)}\approx(5\alpha/2R^2)\ln(R/\xi)\omega_x$,
where $R=\sqrt{2\mu_{\rm TF}}$ and $\xi=\sqrt{\alpha}/R$ are the dimensionless
condensate radius along $\hat{x}$ and the healing length,
respectively~\cite{Fetter1,Feder1}. For the parameters considered here, the
values are predicted to be $\Omega_c^{(1)}=\{0.61,\, 0.33,\,0.16\}\omega_x$
and are found numerically to be
$\Omega_c^{(1)}=\{0.54,\, 0.29,\,0.14\}\omega_x$.
The number of vortices $n_{\nu}$ present just above $\Omega_{\nu}$ is found to
increase with $N_0$; $n_{\nu}=4$ and $8$ for $N_0=10^6$ and $10^7$,
respectively. The value of $\Omega_{\nu}$ may be interpreted as the critical
frequency $\Omega_c^{(n_{\nu})}$ for the stabilization of $n_{\nu}$ vortices.
If $n_{\nu}=n$ for all $N_0$, then
$\Omega_{\nu}\sim\Omega_c^{(n)}\sim N_0^{-2/5}$. That $\Omega_{\nu}$ decreases
more slowly with $N_0$ implies that $n_{\nu}$ must increase with $N_0$. The
small difference between $\Omega_c^{(1)}$ and $\Omega_{\nu}$ for $N_0=10^5$
reflects the instability of vortex arrays in the low-density limit. As $N_0$
decreases, the spacing between successive $\Omega_c^{(n)}$ diminishes, and
vanishes for $N_0=0$ in cylindrically-symmetric traps; for very large $N_0$,
the spacing approaches a constant as the vortex-vortex interactions become
negligible.

The numerical results for $\Omega_{\nu}$ suggest that the criteria for vortex
stabilization and nucleation are different. Superflow through microapertures,
or the motion of an object or ion through a superfluid, can give rise to
vortex half-ring~\cite{Burkhart,Packard3} or
vortex-pair~\cite{McCann,Adams,Davies,Frisch,Ketterle2} production through the
accumulation of phase-slip. One might expect similar excitations in a rotating
condensate~\cite{Jones,Kusmartsev}: vortex half-rings would be nucleated at
the condensate surface when the local tangential velocity exceeds a critical
value. Indeed, the distinction between a half-ring and vortex becomes blurred
in a trapped gas with curved surfaces, as discussed further below.

A crude estimate of $\Omega_{\nu}$ may be obtained by invoking the Landau
criterion for the critical velocity $v_{\rm cr}={\rm min}(\omega_q/q)$, where
$\omega_q$ is the frequency of the mode at wavevector $q$. Such a minimum
corresponds to values of $q_c$ at which the hydrodynamic description of the
collective excitations begins to fail~\cite{Dalfovo}. For a spherical trapped
Bose gas, the crossover to a single-particle behavior occurs in a
boundary-layer region at the cloud surface whose thickness is several
$\delta=(2R)^{-1/3}d_x$~\cite{Pethick,Feder3}. Minimizing $\omega_q/q$ using
the dispersion relation for the planar surface modes~\cite{AlKhawaja} of such a
system
$\omega_q^2\approx\omega_x^2[qR-d_x^4q^4\left(\ln(q\delta)-0.15\right)]$, one
obtains $q_c=(R/0.3)^{1/3}d_x^{-1}\approx\delta^{-1}$ and
$\Omega_{\nu}=v_{\rm cr}/R\sim R^{-2/3}$. Since $R\sim N_0^{1/5}$, the
critical frequency $\Omega_{\nu}\sim N_0^{-2/15}$ decreases far more slowly
than does the TF estimate for $\Omega_c\sim N_0^{-2/5}$. The number-dependence
of $\Omega_{\nu}$ is in reasonable agreement with the numerical data.
Real-time simulations further confirm that high-frequency oscillations of the
condensate are required for vortex production at the same $\Omega_{\nu}$
found using the imaginary-time approach.

The above analysis does not clearly identify the instability of the surface
modes with the penetration of vortices into the cloud, however. Further
insight may be gained by considering the free energy $F$ of a single vortex in
a cylindrical trap, relative to that of the vortex-free state, as a function
of the vortex displacement $\rho$ from the trap center~\cite{Fetter1}. In the
TF limit, $F$ vanishes for $\rho^2=R^2$ and $\rho^2=R^2-(5/2\Omega)\ln(R/\xi)$,
corresponding to the right and left roots of the free energy barrier to vortex
generation, respectively. As $\Omega$ increases, the energy barrier at the
surface narrows but remains finite. Yet, as discussed above, the hydrodynamic
excitations begin to break down at a radius $\tilde{\rho}\approx R-\delta$.
Vortices will therefore spontaneously penetrate the cloud when the angular
frequency exceeds 
\begin{equation}
\Omega_{\nu}={5{\root 3\of 2}\over 4R^{2/3}}\ln\left({R\over\xi}\right)\omega_x,
\end{equation}
since the barrier effectively disappears when $\tilde{\rho}\leq\rho$. Thus,
the frequencies of nucleation and penetration have the same number-dependence
and are defined by a single critical wavelength. Once the condensate contains
vortices at a given $\Omega>\Omega_{\nu}$, the functional $F$ will again
include a barrier to vortex penetration from the surface, reflecting the
hydrodynamic stability of the vortex state. One may thus envisage a succession
of multiple-vortex nucleation events at well-defined angular frequencies.

The stationary configurations of vortex arrays are shown as a function of
applied rotation in Fig.~\ref{rot}. The condensate density is shown integrated
down the axis of rotation $\hat{z}$, in order to mimic an {\it in situ} image
of the cloud. While the vortices near the origin appear to have virtually
isotropic cores, those in the vicinity of the surface are generally wider and
are noticeably distorted. The anisotropy is due in part to the divergence of
the coherence length as the density decreases, but is mostly the result of
vortex curvature. Off-center vortices are not fully aligned with the axis of
rotation $\hat{z}$, since they terminate at normals to the ellipsoidal
condensate surface. Far from the origin, the vortex structure approaches that
of a half-ring pinned to the condensate surface.

The symmetries of the confining potential impose constraints on the vortex
arrays that may be produced by rotating anisotropic traps. Stationary
configurations are found to always have the inversion symmetry
$(x,y,z)\to(-x,-y,z)$. As shown in Fig.~\ref{rot}, the number of vortices is
at least four and is even for each array; real-time simulations demonstrate
that vortices with the same circulation are nucleated in pairs at
inversion-related points on the surface. No vortex is found at the origin,
since the odd number of remaining vortices cannot be distributed symmetrically.
At low angular velocities, therefore, the array tends to approximate a regular
tetragonal lattice. As the total number of vortices increases with $\Omega$,
however, a different pattern begins to emerge. While a triangular array is
inconsistent with the twofold trap symmetries, it is more efficient for close
packing; this geometry is favored for vortices near the rotation axis of
rapidly rotating vessels of superfluid
helium~\cite{Packard1,Tkachenko,Campbell}. If vortices in trapped condensates
could be made sufficiently numerous, they would likely form a near-regular
triangular array but with the central vortex absent.

In summary, the critical frequencies for the zero-temperature nucleation of
vortices $\Omega_{\nu}$ in rotating anisotropic traps are obtained numerically,
and are found to be larger than the vortex stability frequencies $\Omega_c$.
The number-dependence of $\Omega_{\nu}$ is consistent with a critical-velocity
mechanism for vortex production. The structures of vortex arrays are strongly
affected by trap geometry, but approach triangular at large densities.

\begin{acknowledgments}
The authors are grateful to A.~L.~Fetter, R.~L.~Pego, S.~L.~Rolston,
J.~Simsarian, and S.~Stringari for numerous fruitful discussions, and to
P.~Ketcham for assistance in generating the figures. This work was supported
by the U.S.\ office of Naval Research.
\end{acknowledgments}

\newpage

\begin{table}[t]
\begin{center}
\begin{tabular}{lccc}
$n$ & $\mu$ ($\hbar\omega_x$) & $E$ ($\hbar\omega_x$)
& $\langle L_z\rangle$ ($\hbar$) \cr
\hline
$0$ & $19.874$ & $14.339$ & $0.779$ \cr
$1$ & $19.758$ & $14.196$ & $1.611$ \cr
$2$ & $19.624$ & $14.139$ & $2.355$ \cr
$3$ & $19.553$ & $14.130$ & $2.864$ \cr
$4^*$ & $19.517$ & $14.134$ & $3.157$ \cr
\end{tabular}
\end{center}
\caption{The chemical potential $\mu$, ground state energy $E$ and average
angular momentum per particle $\langle L_z\rangle$ are given as a function of
the number of vortices $n$ for $N_0=10^6$ and applied angular frequency
$\Omega=0.45\omega_x$. The $n=4$ solution may be metastable since $\mu_4<\mu_3$
while $E_4>E_3$.}
\label{tablerot}
\end{table}

\end{document}